\documentclass[twocolumn]{aastex631}
\newcommand{\g}{G296.1$-$0.5}
\newcommand{\fr}{$f/r$ ratio}
\newcommand{\frs}{$f/r$ ratios}
\newcommand{\oseven}{\ion{O}{7}}
\received{February 16, 2022}
\revised{April 19, 2022}
\accepted{May 19, 2022}
\shorttitle{Charge Exchange X-ray Emission in SNR G296.1$-$0.5}
\shortauthors{Tanaka et al.}

\begin{document}

\title{Charge Exchange X-ray Emission Detected in Multiple Shells of Supernova Remnant G296.1$-$0.5}


\author[0000-0003-1857-7425]{Yukiko Tanaka}
\affiliation{Department of Physics, Kyoto University,
Kitashirakawa Oiwake-cho, Sakyo, Kyoto,
Kyoto 606-8502, Japan}
\email{tanaka.yukiko.x13@kyoto-u.jp}

\author[0000-0003-1518-2188]{Hiroyuki Uchida}
\affiliation{Department of Physics, Kyoto University,
Kitashirakawa Oiwake-cho, Sakyo, Kyoto,
Kyoto 606-8502, Japan}

\author[0000-0002-4383-0368]{Takaaki Tanaka}
\affiliation{Department of Physics, Konan University,
8-9-1 Okamoto, Higashinada, Kobe,
Hyogo 658-8501, Japan}
\affiliation{Department of Physics, Kyoto University,
Kitashirakawa Oiwake-cho, Sakyo, Kyoto,
Kyoto 606-8502, Japan}

\author[0000-0003-4520-9505]{Yuki Amano}
\affiliation{Department of Physics, Kyoto University,
Kitashirakawa Oiwake-cho, Sakyo, Kyoto,
Kyoto 606-8502, Japan}

\author[0000-0002-3119-2928]{Yosuke Koshiba}
\affiliation{Department of Physics, Kyoto University,
Kitashirakawa Oiwake-cho, Sakyo, Kyoto,
Kyoto 606-8502, Japan}

\author[0000-0002-5504-4903]{Takeshi Go Tsuru}
\affiliation{Department of Physics, Kyoto University,
Kitashirakawa Oiwake-cho, Sakyo, Kyoto,
Kyoto 606-8502, Japan}

\author[0000-0003-2062-5692]{Hidetoshi Sano}
\affiliation{Faculty of Engineering, Gifu University,
1-1 Yanagido,
Gifu 501-1193, Japan}

\author[0000-0002-8966-9856]{Yasuo Fukui}
\affiliation{Department of Physics, Nagoya University,
Furo-cho, Nagoya,
Aichi 464-8601, Japan}

\begin{abstract}
Recent high-resolution X-ray spectroscopy revealed possible presence of charge exchange (CX) X-ray emission in supernova remnants (SNRs).
Although CX is expected to take place at outermost edges of SNR shells, no significant measurement has been reported so far due to the lack of  nearby SNR samples.
Here we present an X-ray  study of SNR \g, which has a complicated multiple-shell structure, with the Reflection Grating Spectrometer (RGS) onboard XMM-Newton.
We select two shells in different regions and find that in both regions \oseven \ line shows a high forbidden-to-resonance ($f/r$) ratio that cannot be reproduced by a simple thermal model.
Our spectral analysis suggests a presence of CX and the result is also supported by our new radio observation, where we discover evidence of  molecular clouds associated with these shells.
Assuming \g \ has  a spherical  shock, we estimate that CX is dominant in a thin layer with a thickness of 0.2--0.3\% of the shock radius.
The result is consistent with a previous theoretical expectation and we therefore conclude that CX occurs in \g.
\end{abstract}


\section{Introduction} \label{sec:intro}
High-resolution X-ray spectroscopy is one of the key methods to investigate plasma conditions and radiation mechanisms of astronomical objects.
Plasma diagnostics for supernova remnants (SNRs) provides us with important clues to understand interactions between expanding shocks and ambient media.  
In this context, previous studies were performed  mainly with the Reflection Grating Spectrometer \citep[RGS;][]{herder2001} onboard XMM-Newton; some of them revealed anomalously high forbidden-to-resonance intensity ratios (\frs) of  the \oseven \ He$\alpha$ line \citep[e.g.,][]{heyden2003, broersen2011}.
Although their physical origin is still unclear, recent studies indicate that charge exchange (CX) is one of the possibilities to explain the high \fr \ \citep[e.g.,][]{katsuda2012, uchida2019, koshiba2022}.

CX takes place when highly charged ions collide with a cold neutral matter population.
Thus, indicating that SNR-cloud interaction regions are plausible candidates to detect CX X-ray emission.
Since the CX cross section for ions of light elements (e.g., O$^{8+}$) peaks at a few thousand km~s$^{-1}$ \citep[e.g.,][]{gu2016}, evolved SNRs are the best targets to search for CX X-ray emissions.
Possible detections are indeed reported from observations of middle-aged SNRs such as Puppis~A \citep{katsuda2012} and the Cygnus Loop \citep{uchida2019}.
If CX is occurring in these SNRs, its emitting region would be the outer edge of the shock front (a few percent of the shock radius) in contact with dense clouds \citep{lallement2004}.
However, a quantitative measurement together with direct evidence for cloud interactions are both missing so far in previous studies.

\g \ is a nearby middle-aged \citep[$\sim$ 20,000~yr;][]{hwang1994} SNR with a diameter of 35-50~pc \citep[assuming a distance of 4~kpc][]{longmore1977}.
Previous radio and X-ray observations revealed a complex asymmetric morphology, especially a double-shell structure in the south \citep{merkert1981, whiteoak1996}, which implies a dense inhomogeneous medium  in the vicinity of the remnant.
This is supported by \citet{castro2011}, who performed X-ray imaging and spectral analysis of \g \ with XMM-Newton and concluded that this remnant is of Type Ib/c origin evolved in a dense environment.
The presence of the dense ambient medium is suggested also by H$\alpha$ emissions detected at several locations over \g \ \citep{gaustad2001}.
These characteristics are common with Puppis~A and the Cygnus Loop, and thus \g \ can be another potential candidate for a CX-dominant SNR.

In this paper, we perform a high-resolution spectroscopy for two shells of \g \ using the RGS, mainly focusing on detection of  the CX X-ray emissions.
We also carry out a $^{12}$CO($J$ = 1--0) and \ion{H}{1} observation to probe molecular clouds and atomic gas around the remnant.
These observations enable us to investigate physical conditions that enhance the contribution of CX in SNRs.
Throughout the paper, errors are given at the 68\% confidence level.

\section{Observations and Data Reduction} \label{sec:obs}
\g \ was observed four times with XMM-Newton from 2007 to 2010. 
Each observation targets different parts of the SNR so as to cover the whole remnant.
Among them, one of the observations of the southwestern region was contaminated by a stellar flare occurred in the field of view (FOV).
We therefore used only another dataset  (Obs.ID~0503220201) for the southwest.
Those of the northwestern (Obs.ID~0503220101) and eastern (Obs.ID~0503220301) shell  regions were also used for the following imaging analysis.
To perform a high-resolution spectroscopy, we selected two datasets (northwest and southwest; hereafter, NW and SE), in which the shells are along the cross-dispersion direction in the RGS FOV, since the energy resolution of the RGS highly depends on the source width.
A nearby blank-sky observation (Obs.ID~0804240201) outside the SNR was used to estimate the background. 

\begin{deluxetable*}{ccccc}
\tablenum{1}
\tablecaption{Observation Data\label{tab:obs}}
\tablewidth{0pt}
\tablehead{
\colhead{Region} & \colhead{Observation ID} & \colhead{Starting Time} & \multicolumn{2}{c}{Effective Exposure Time (ks)}  \\
    &   &  & \colhead{MOS} & \colhead{RGS} 
}
\startdata
NW & 0503220101 & 2007 Jul 7 & 22.2 & 25.4\\
SE & 0503220201 & 2007 Jul 15 & 4.4 & 25.8\\
East & 0503220301 & 2007 Dec 24  & 7.7 & --- \\
Background & 0804240201 & 2018 Feb 8 & 94.0 & 117.3\\
\enddata
\tablecomments{The effective time of MOS represents the sum of MOS~1 and MOS~2. The same is true for RGS.}
\end{deluxetable*}

\begin{figure}[t!]
\centering
\includegraphics[clip,width=90 mm]{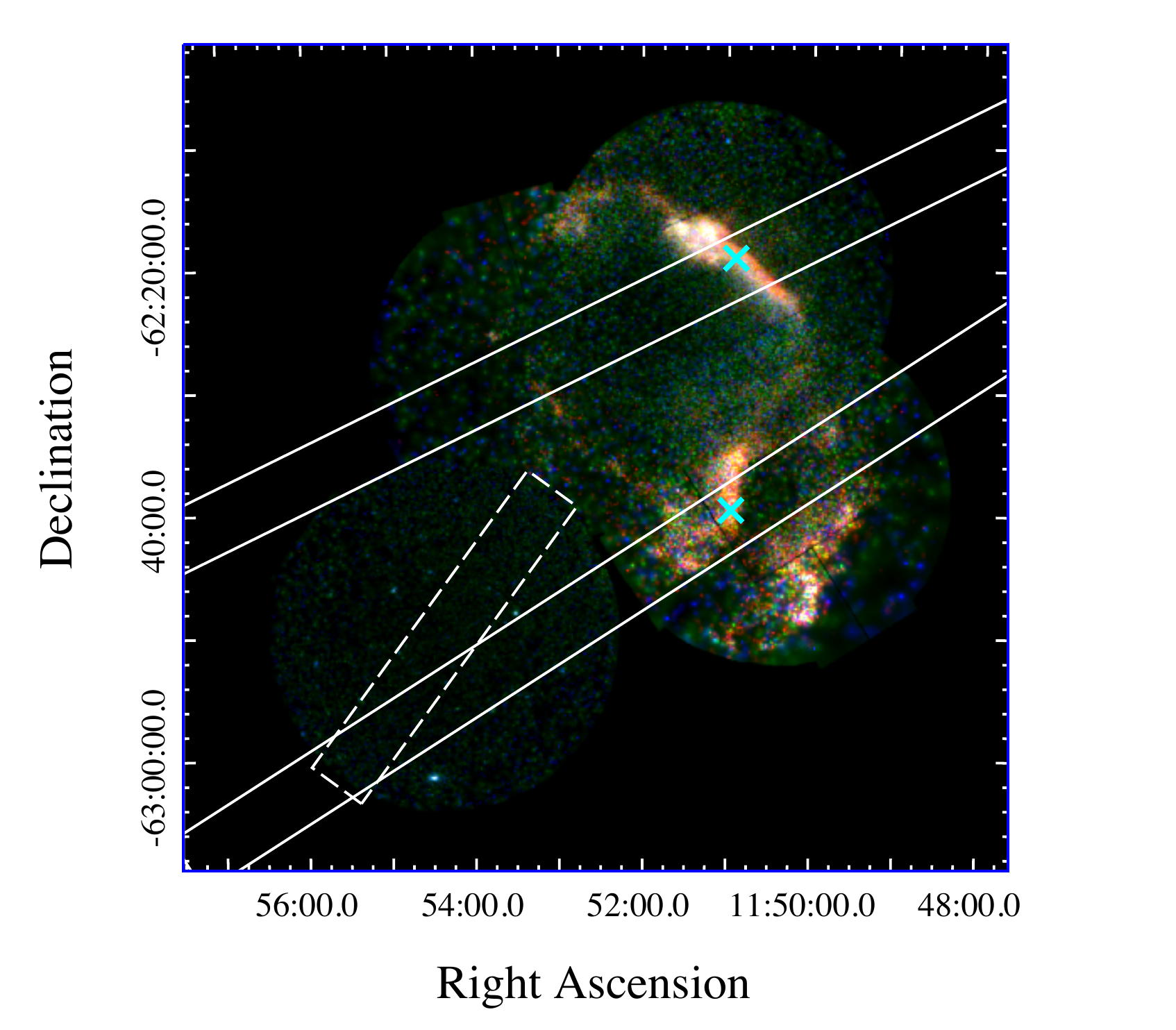}
\caption{Three-color image of G296.1$-$0.5. 
Red, green and blue correspond to the 0.3--0.7~keV, 0.7--1.0~keV, and 1.0--1.5~keV energy bands.
The white lines represents the cross-dispersion widths of the RGS (5\arcsec).  
The white dotted rectangle shows the region used as a background.
The cyan X marks are the center of each source position; ($\alpha_{2000}$, $\delta_{2000}$) = ($11^{h}50^{m}54.5059^{s}$, $-62^\circ19\arcmin03.44\arcsec$), ($11^{h}50^{m}57.2627^{s}$, $-62^\circ38\arcmin46.40\arcsec$) for NW and SE, respectively.
}
\label{fig:obs}
\end{figure}

We used XMM Science Analysis Software (SAS) version 18.0.0 for the following analysis.
The European Photon Imaging Camera MOS \citep[EPIC MOS;][]{turner2001} data were processed using the XMM-Newton Extended Source Analysis Software package (XMM-ESAS) with background processing based on the modeling of \citet{snowden2004}.
Figure~\ref{fig:obs} shows a vignetting-corrected three-color image of \g.
We set the brightest locations of the NW and SE shells as the ``source positions'' (Figure~\ref{fig:obs}) and processed the RGS data using the standard pipeline tool \texttt{rgsproc}.
The good time interval was determined based on the count rate of CCD9, which is closest to the optical axis of the telescope and most affected by background flares.

As a result of the above procedure, we obtained effective times for each instrument as summarized in Table~\ref{tab:obs}.
Note that we used only first-order spectra since second-order spectra do not have sufficient statistics.
Model fitting was performed in the range of 8.3--31~\AA \  (0.40--1.5~keV) since backgrounds dominate outside the range. 

\section{Spectral Analysis} \label{sec:ana}
\begin{figure*}[th]
\centering
\includegraphics[width=150 mm]{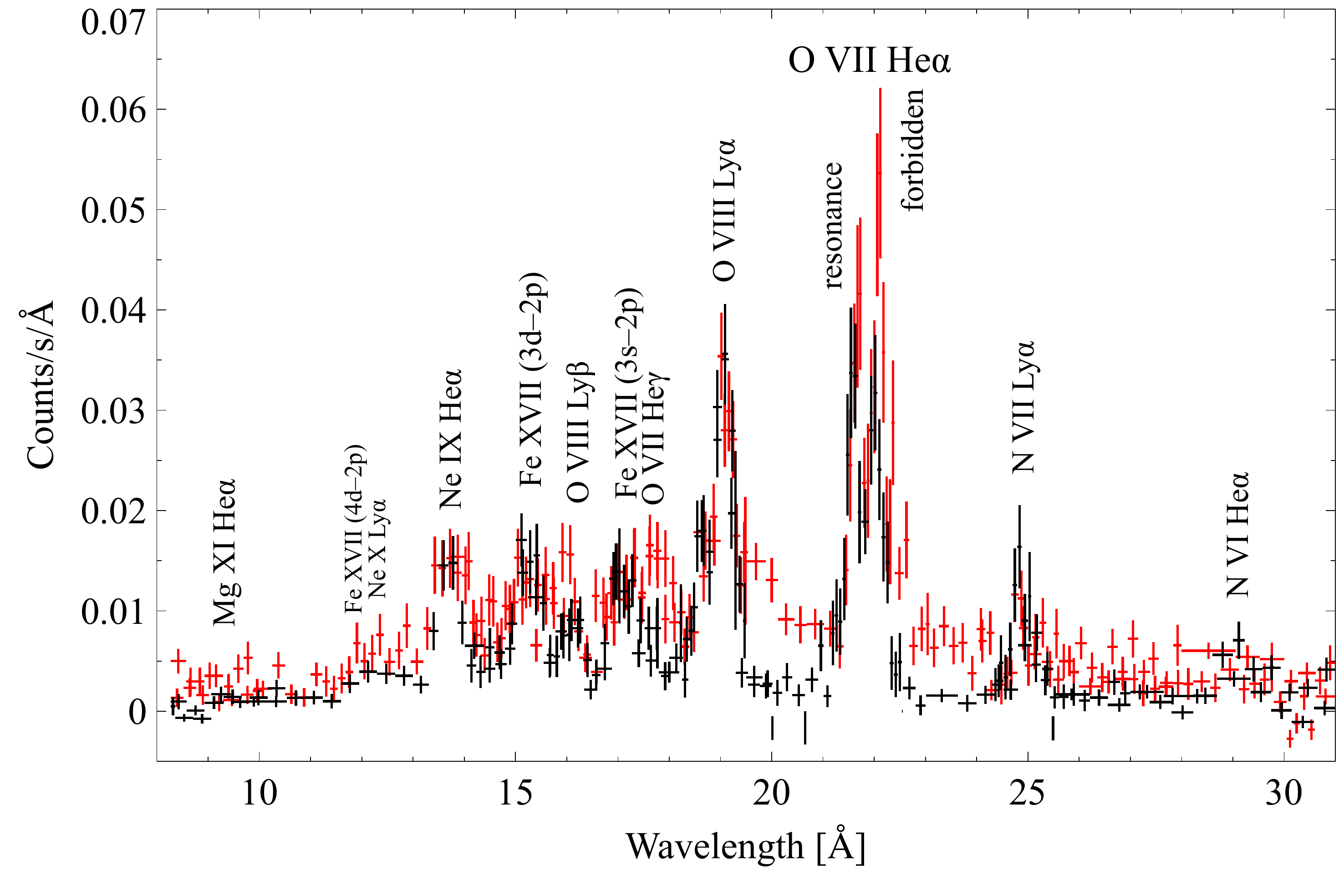}
\caption{RGS spectra of the NW shell (black) and the SE shell (red).}
\label{allsp}
\end{figure*}

The RGS spectra extracted from the NW and SE shells are displayed in Figure~\ref{allsp}.
Clearly visible in the spectra are the \ion{Mg}{11} He$\alpha$ (9.2~\AA), \ion{Ne}{9} He$\alpha$ (13.5~\AA), \ion{Fe}{17} (15.0~\AA; 17.0~\AA), \ion{O}{8} Ly$\alpha$ (19.0~\AA), \oseven \ He$\alpha$ ($\sim22$~\AA), \ion{N}{7} Ly$\alpha$ (24.8~\AA) and \ion{N}{6} He$\alpha$ (28.8~\AA) lines.
We found that the \oseven \ He$\alpha$ line is clearly resolved into the resonance and forbidden lines and that the forbidden line seems relatively stronger than the resonance line especially in the spectrum of SE.
The following spectral fitting was performed simultaneously for the RGS1/2 and MOS1/2 spectra.
We used SPEX version 3.06.01 \citep{kaastra1996} for the spectral fit with the maximum likelihood W-statistic \citep{wstat1979}. 
To take into account the spatial extent of the source, we multiplied a spectral model with \texttt{Lpro}\footnote{\url{https://spex-xray.github.io/spex-help/models/lpro.html}}, which convolves a model function with a surface brightness profile, including a point-spread function, which is estimated from the MOS1 image \citep{tamura2004}.

We first tried fitting the spectra with the non-equilibrium ionization (NEI) model \texttt{neij} absorbed by neutral gas with the cosmic abundances \citep{morrison1983}.
This ``NEI$\times$abs'' model was used also in the analysis by \citet{castro2011} of the EPIC MOS data.
We set the volume emission measure $\rm{VEM}_{\rm{NEI}}$ $(=n_{\rm{e}} n_{\rm{p}} V_{\mathrm{NEI}}$, where $n_{\rm{e}}$, $n_{\rm{p}}$, and $V_{\mathrm{NEI}}$ are the electron density, proton density, and the volume of NEI plasma, respectively), the electron temperature $kT_{\rm{e}}$, and the ionization timescale $n_{\rm{e}}t$ as variables.
The abundances of N, O (=C), Ne, Mg, Fe (=Ni) were left free and the others were fixed at the solar values estimated by \citet{Lodders2009}. 
The absorption column density $N_{\rm{H}}$ was also a variable.

\begin{figure*}[th]
\centering
\includegraphics[width=85 mm]{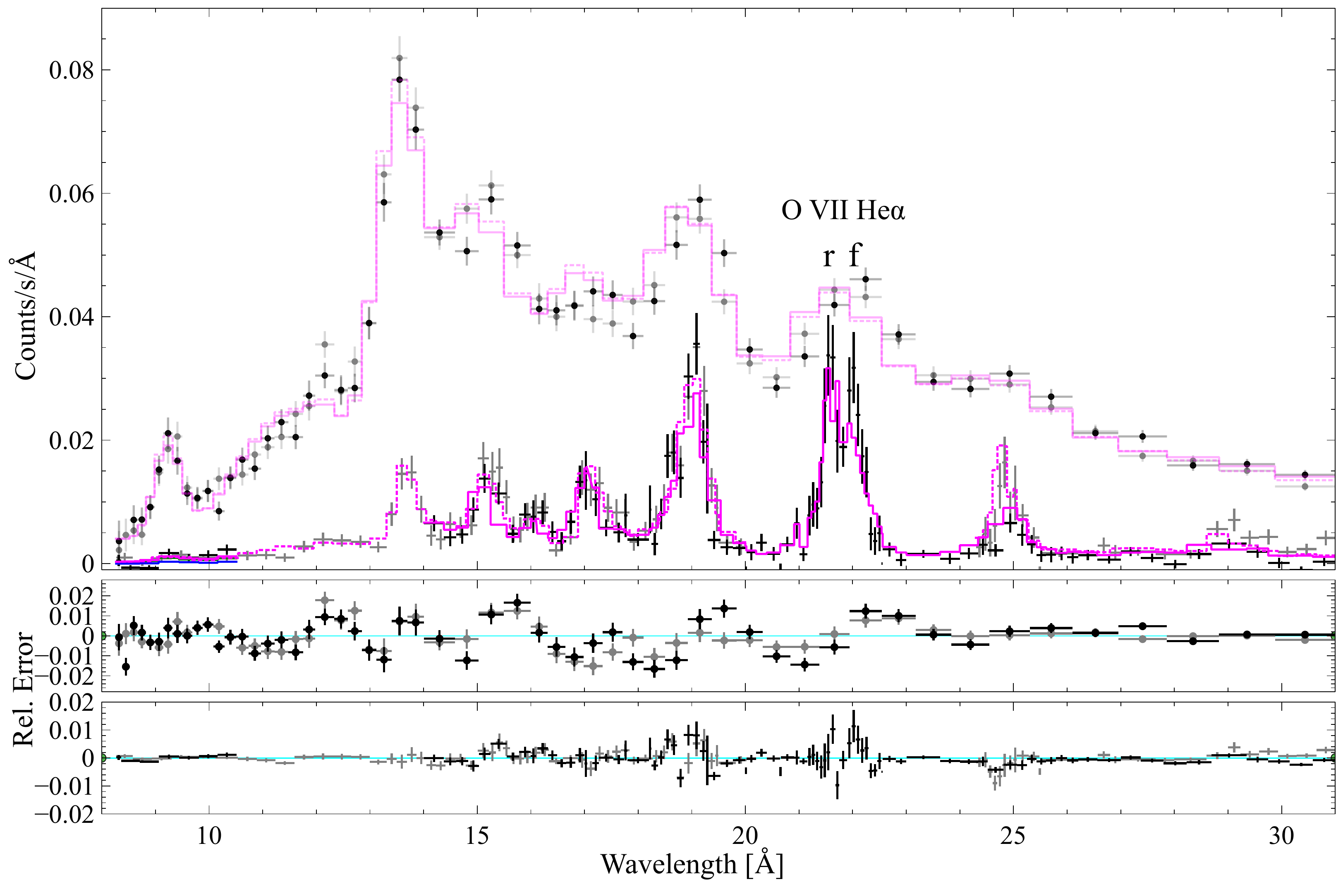}
\includegraphics[width=85 mm]{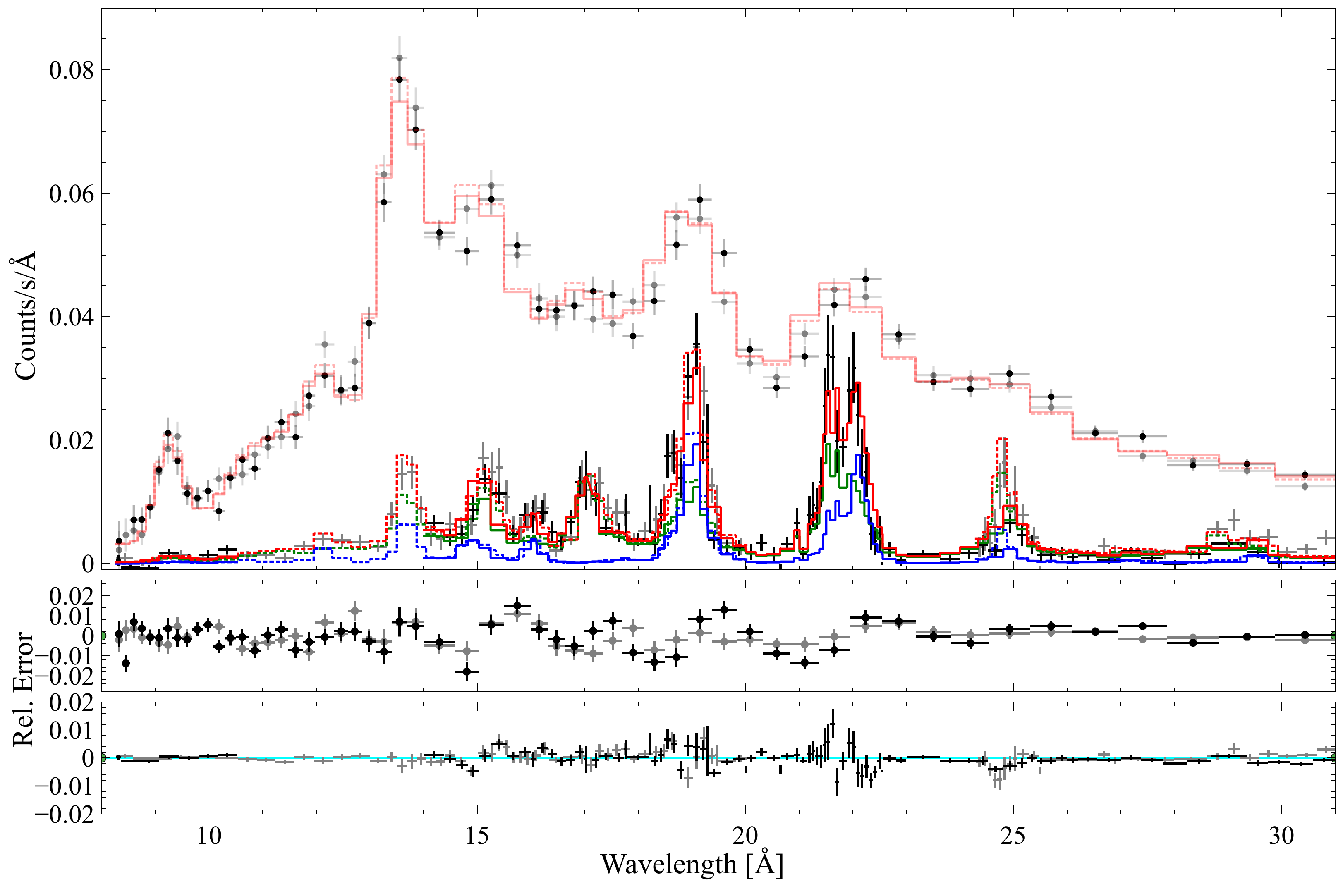}
\includegraphics[width=85 mm]{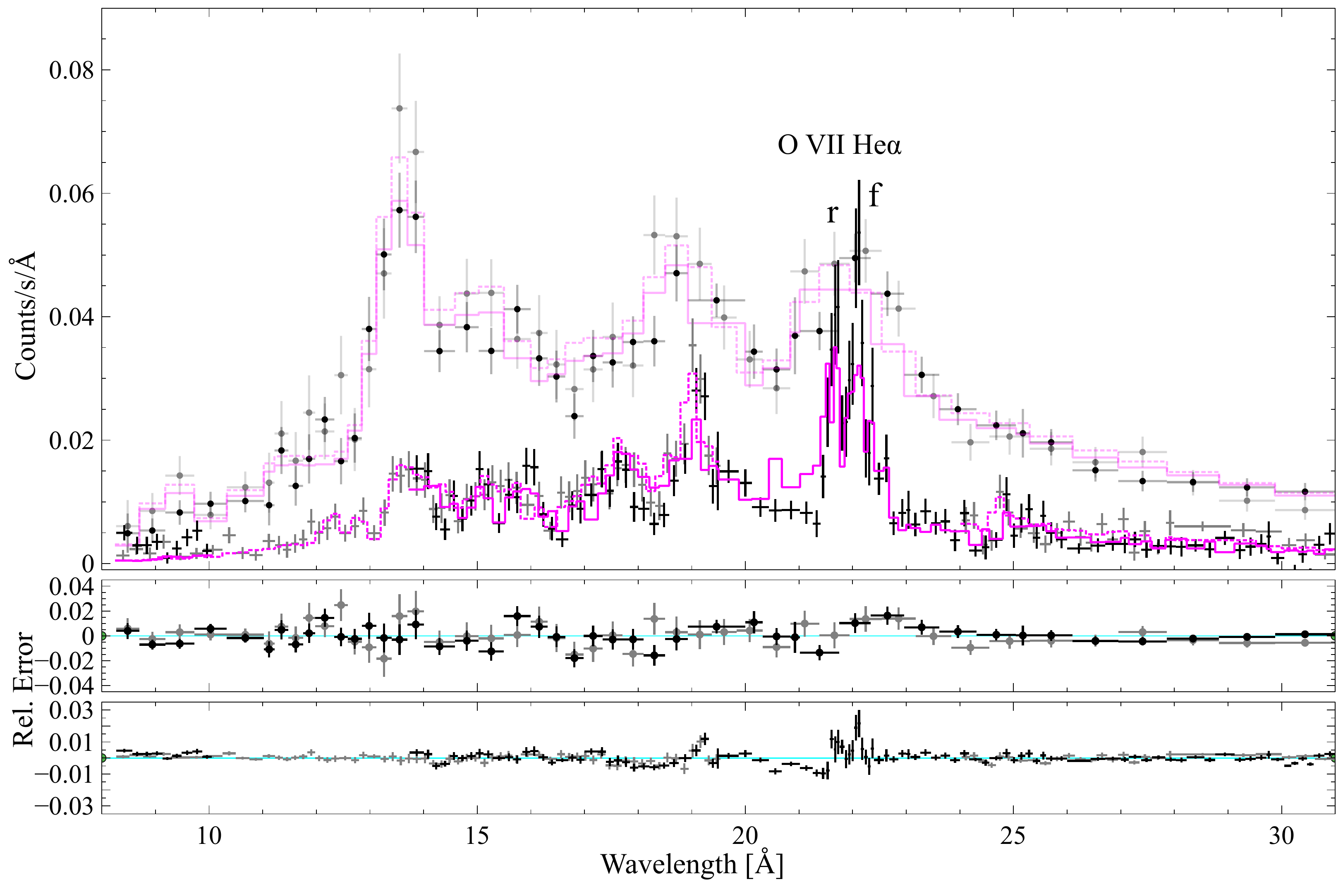}
\includegraphics[width=85 mm]{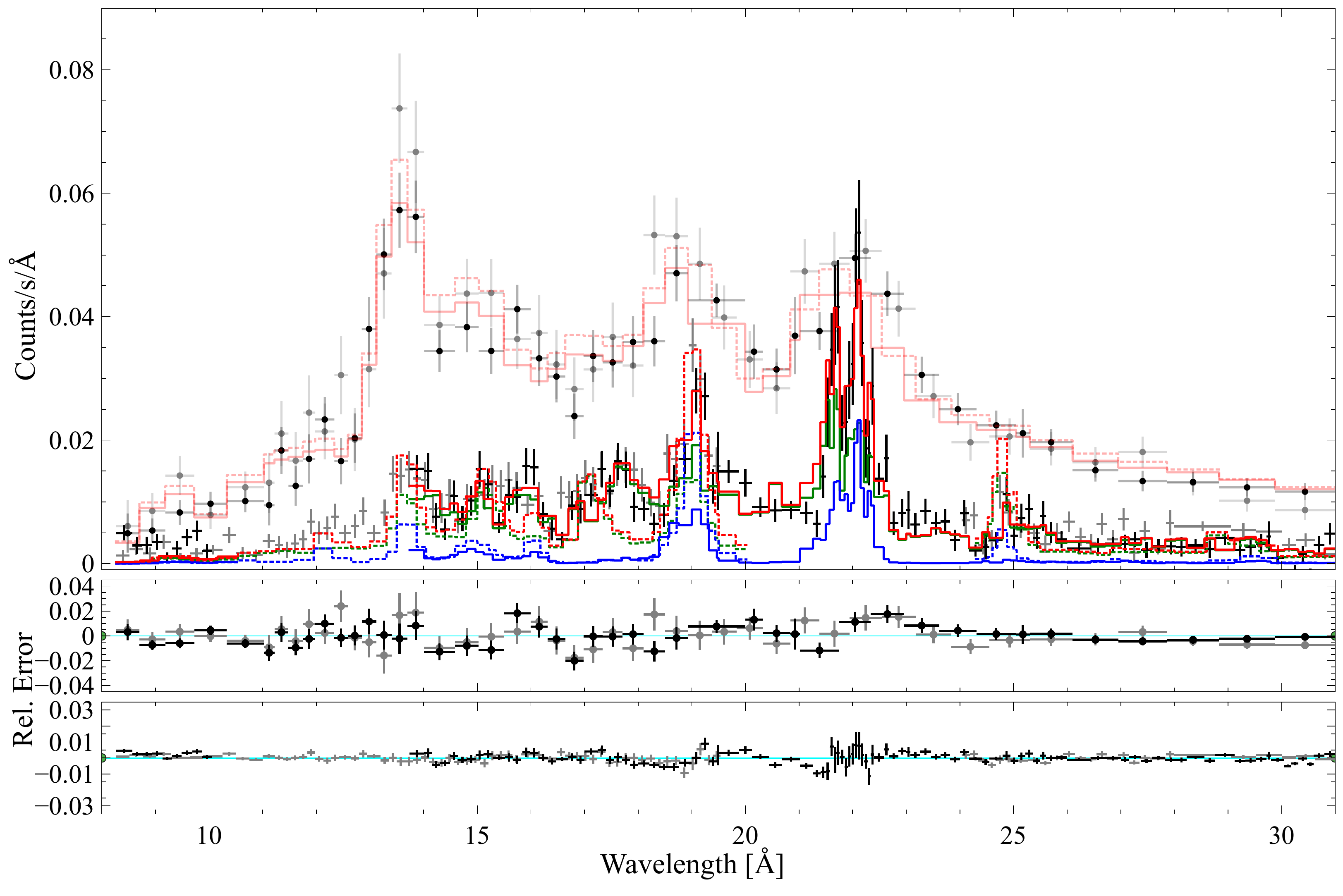}
\caption{\textit{Top}: MOS1 (black circle), MOS2 (grey circle), RGS1 (black cross) and RGS2 (gray cross) spectra of the NW shell. The solid and dotted lines are the best-fit models for MOS1/RGS1 and MOS2/RGS2, respectively. 
The MOS spectra are scaled by a factor of 0.5 for a displaying purpose.
The best-fit results of the NEI$\times$abs and (NEI$+$CX)$\times$abs models are shown in the left and right panels, respectively.
The middle and lower panels in each figure represent the residuals for the MOS and the RGS, respectively.
\textit{Bottom}: Same as the top panels, but for the SE shell.}
\label{nwpl}
\end{figure*}

The best-fit NEI$\times$abs models for the NW and SE shells are shown in Figure~\ref{nwpl}.
The best-fit parameters are listed in Table~\ref{tab:par}. 
There are large residuals at the wavelength of \oseven \ He$\alpha$; the forbidden lines in the NW and SE spectra are significantly stronger than the prediction by the model.
While the result for NW was already implied by the result of \citet{castro2011} (see their Figure~5), we discovered that the SE shell also shows a similar sign of a strong forbidden line.

\begin{deluxetable*}{cccc}
\tablenum{2}
\tablecaption{Best-fit parameters\label{tab:par}}
\tablewidth{0pt}
\tablehead{
\colhead{Region} & \colhead{Parameter} & \colhead{NEI$\times$abs}  & \colhead{(NEI + CX) $\times$ abs}
}

\startdata
NW & $N_{\rm{H}}$ ($10^{20}\mathrm{cm^{-2}}$) & $\leq 2.4$ & $5.3^{+1.8}_{-1.1}$ \\
    & $kT_{\rm{e}}$ (keV) & $0.50^{+0.06}_{-0.02}$ & $0.40^{+0.01}_{-0.01}$ \\
    & $n_{\rm{e}}t$ ($10^{10}\mathrm{s~cm^{-3}}$) & $2.8^{+0.3}_{-0.4}$ & $3.0^{+0.3}_{-0.3}$ \\
        & $v_{\rm{col}}$ (km~s$^{-1}$) & -- & $394^{+72}_{-48}$ \\
    & N & $0.87^{+0.20}_{-0.11}$ & $0.63^{+0.07}_{-0.07}$ \\
    & O(=C) & $0.33^{+0.06}_{-0.03}$ & $0.17^{+0.02}_{-0.02}$ \\
    & Ne & $0.54^{+0.12}_{-0.04}$ & $0.36^{+0.03}_{-0.04}$ \\
    & Mg & $0.41^{+0.10}_{-0.04}$ & $0.38^{+0.05}_{-0.05}$ \\
    & Fe(=Ni) & $0.39^{+0.09}_{-0.04}$ & $0.33^{+0.03}_{-0.03}$ \\
    & $\rm{VEM}_{\rm{NEI}} (10^{56} \mathrm{cm^{-3}})$ & $0.6^{+0.1}_{-0.2}$ & $1.0^{+0.3}_{-0.1}$ \\
    & $\rm{VEM}_{\rm{CX}} (10^{56} \mathrm{cm^{-3}})$ & -- & $1.1^{+0.9}_{-0.2}$ \\
    & W-statistic/dof & 4382/3905 & 4247/3902 \\ \hline
SE  & $N_{\rm{H}}$ ($10^{20}\mathrm{cm^{-2}}$) & $18.9^{+6.9}_{-1.3}$ & $\leq 7.3$ \\
    & $kT_{\rm{e}}$ (keV) & $0.204^{+0.008}_{-0.047}$ & $0.37^{+0.06}_{-0.02}$ \\
    & $n_{\rm{e}}t$ ($10^{10}\mathrm{s~cm^{-3}}$) & $17 \leq$ & $3.8^{+0.8}_{-1.0}$ \\
    & $v_{\rm{col}}$ (km~s$^{-1}$) & -- & $\leq 213$ \\
    & N & $0.12^{+0.04}_{-0.05}$ & $0.32^{+0.14}_{-0.5}$ \\
    & O(=C) & $0.17^{+0.02}_{-0.03}$ & $0.20^{+0.05}_{-0.02}$ \\
    & Ne & $0.24^{+0.05}_{-0.05}$ & $0.31^{+0.09}_{-0.03}$ \\
    & Mg & $0.35^{+0.10}_{-0.09}$ & $0.28^{+0.11}_{-0.07}$ \\
    & Fe(=Ni) & $0.09^{+0.02}_{-0.02}$ & $0.20^{+0.10}_{-0.02}$ \\
    & $\rm{VEM}_{\rm{NEI}} (10^{56} \mathrm{cm^{-3}})$ & $35^{+65}_{-7}$ & $2.2^{+0.5}_{-1.0}$ \\
    & $\rm{VEM}_{\rm{CX}} (10^{56} \mathrm{cm^{-3}})$ & -- & $1.7^{+0.7}_{-1.0}$ \\
    & W-statistic/dof & 4448/3872 & 4398/3869 \\
\enddata
\tablecomments{The abundances are calculated relative to the proto-Solar abundances of \citet{Lodders2009}.
}
\end{deluxetable*}

\begin{figure}[th!]
\centering
\includegraphics[clip,width=70 mm]{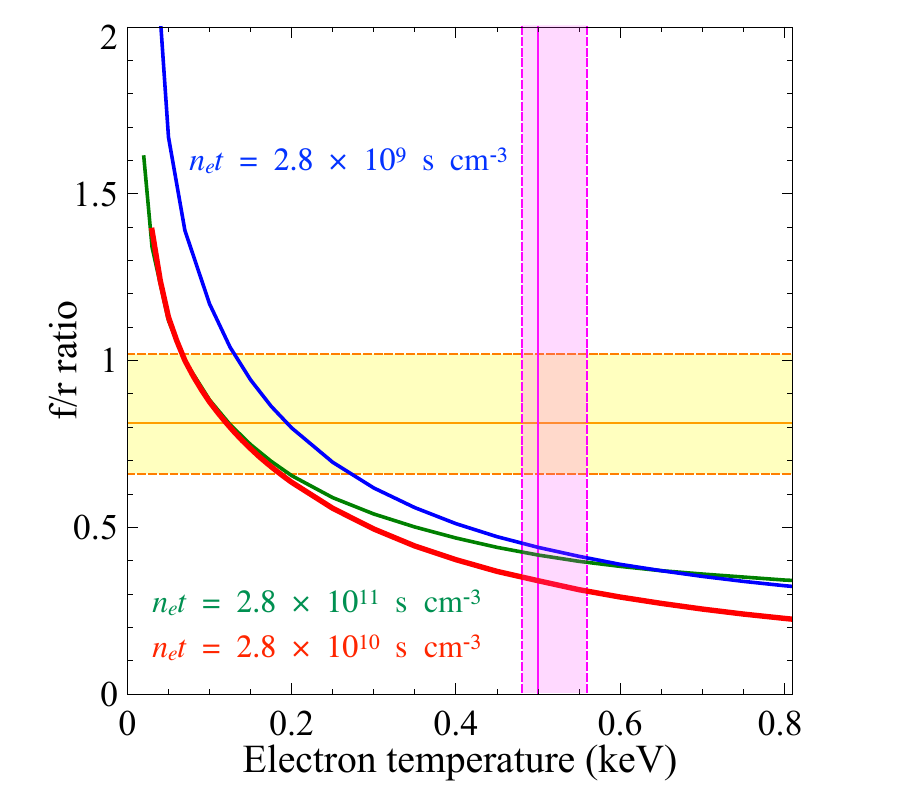}
\centering
\includegraphics[clip,width=70 mm]{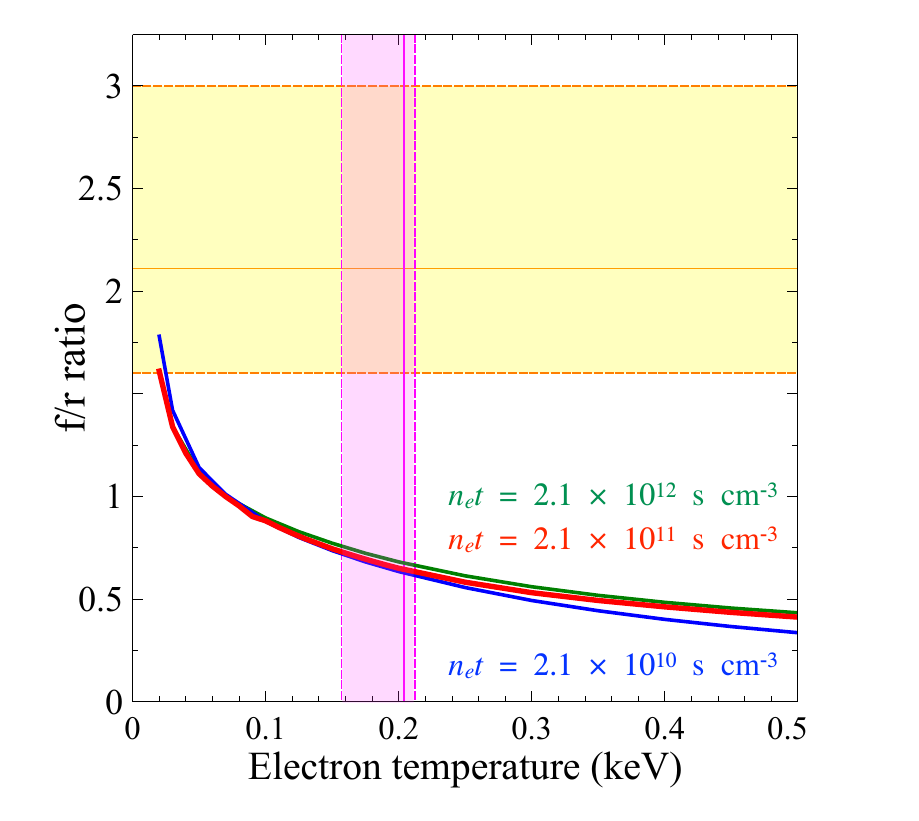}
\caption{\fr \ expected from an NEI model as a function of $kT_{\rm{e}}$.
The top and bottom panels are for the NW and SE shells, respectively.
The colors of each line indicate $n_{e}t$ assumed: the red line corresponds to the best-fit value.
The orange and magenta hatched areas indicate the obtained \fr \ and  the best-fit $kT_{e}$, respectively.}
\label{kevfrratio}
\end{figure}

For more quantitative discussion on the \frs, we calculated the ratios by replacing the \oseven \ lines in the best-fit model with Gaussians, and obtained $0.81^{+0.21}_{-0.15}$ and $2.1^{+0.9}_{-0.5}$ for the NW and SE shells, respectively (see Table~\ref{tab:ratio}).
Figure~\ref{kevfrratio} shows \frs \ expected from the NEI model as a function of $kT_{\rm{e}}$ in the range of possible ionization states $n_{\rm{e}} t$. 
We found that the observed \fr \ for NW requires an extremely low temperature ($<0.2$~keV) if we assume a simple NEI model.
The result is inconsistent with the best-fit value of $kT_{\rm{e}}=0.48$--0.56~keV and, furthermore, such a low-temperature plasma cannot emit detectable \oseven \ lines  due to insufficient amount of O$^{6+}$ ions; the same is true also for the SE shell.
Therefore, another physical process is needed to explain the high \frs \  in \g.

\begin{deluxetable}{cccc}
\tablenum{3}
\tablecaption{{\it f/r} ratios calculated from each model and data\label{tab:ratio}}
\tablewidth{0pt}
\tablehead{
\colhead{Region} & \colhead{Model} & \colhead{component} & \colhead{\fr}
}

\startdata
NW & NEI$\times$abs & NEI  & 0.34\\
    & (NEI + CX) $\times$ abs & NEI & 0.40\\
    &   & CX & 2.2\\ \hline
    & Data &    & $0.81^{+0.21}_{-0.15}$ \\ \hline
SE & NEI$\times$abs & NEI & 0.64 \\
    & (NEI + CX) $\times$ abs & NEI & 0.43\\
    &  & CX & 3.1 \\ \hline
    & Data &    & $2.1^{+0.9}_{-0.5}$ \\ \hline
\enddata
\end{deluxetable}

\begin{figure}[th]
\centering
\includegraphics[clip,width=75 mm]{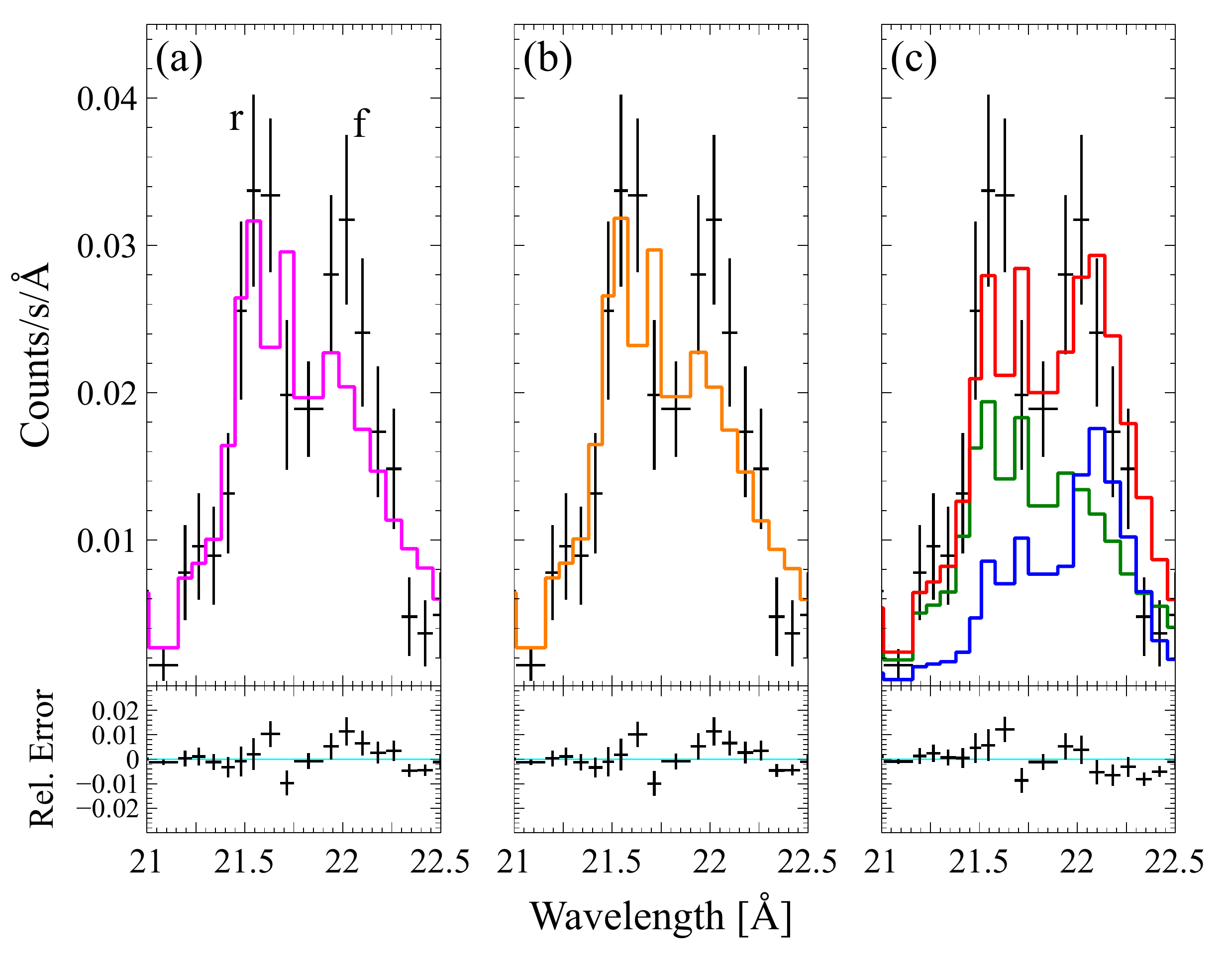}
\centering
\includegraphics[clip,width=75 mm]{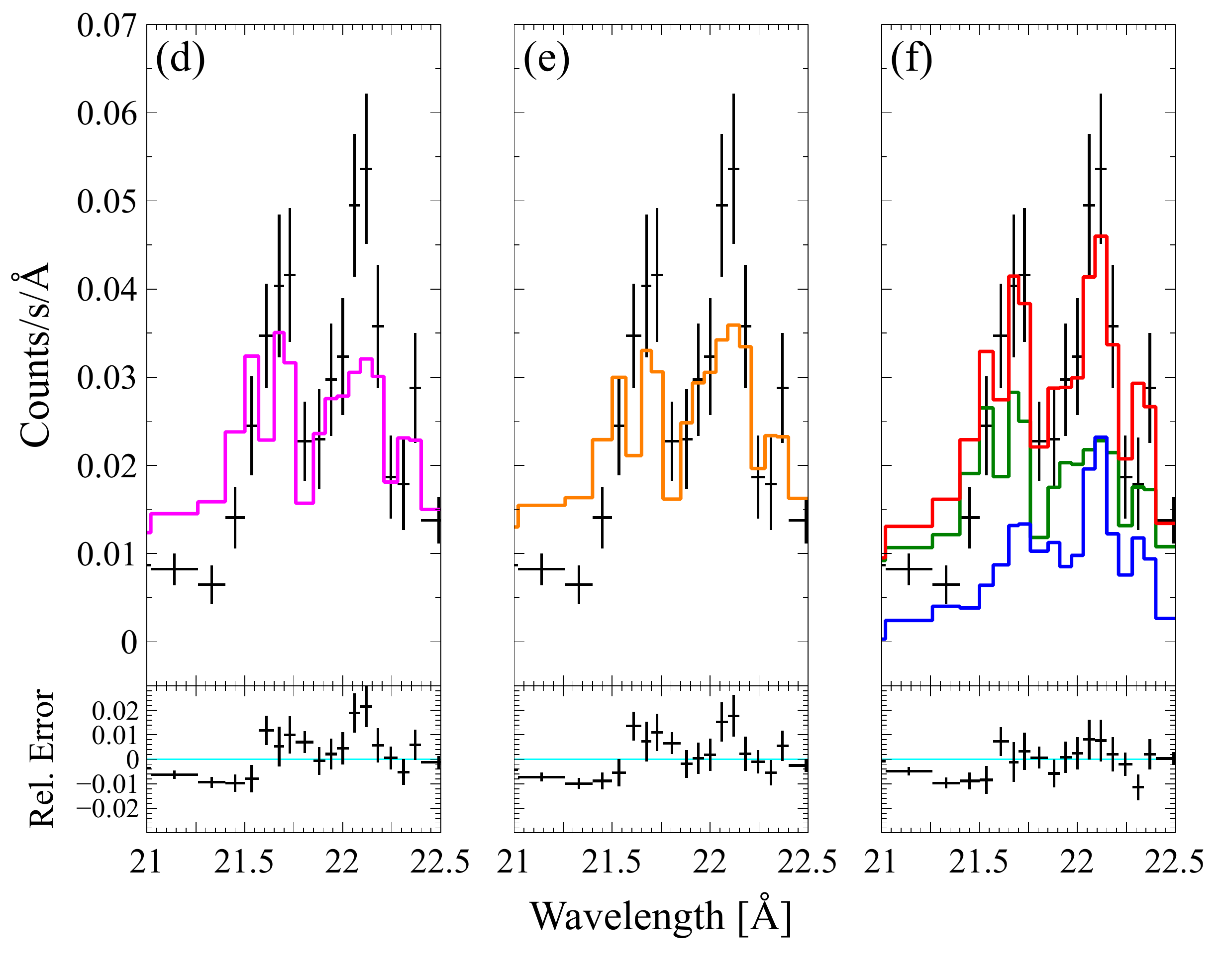}
\caption{Close-up view of the best-fit results around the \ion{O}{7} line for the NW (a--c) and SE (d--f) shells. The magenta, orange, and red lines denote the NEI$\times$abs, (NEI$-$Gaus)$\times$abs, and (NEI$+$CX)$\times$abs  models, respectively. The NEI and CX components are shown as the green and blue lines, respectively. The bottom panels show residuals from the models.}
\label{frcompare}
\end{figure}

Resonance scattering (RS) is one of the possibilities that can enhance the \fr.
RS is considered to be effective for an SNR with an asymmetric morphology and optically thick plasma depth along the line of sight \citep{kaastra1995}.
Clear observational evidence for RS was first discovered by \citet{amano2020}, who analyzed RGS spectra of the extra-Galactic SNR N49.
Following their method to evaluate a contribution of the RS effect,  we tried the ``(NEI$-$Gaus)$\times$abs'' model, in which negative Gaussians were added to the NEI model.
The four Gaussians we added represent the scattered photon fluxes of the resonance lines  of \ion{O}{7} He$\alpha$ (21.6~\AA) and  \ion{Ne}{9} He$\alpha$ (13.4~\AA), \ion{Fe}{17} (3d--2p) (15.0~\AA), and \ion{O}{8} Ly$\alpha$ (19.0~\AA), which have high oscillator strengths.
From the best-fit results, the values of W-statistic/d.o.f. are 4256/3901 and 4411/3868 for NW and SE, respectively.
While a possibility of RS cannot be ruled out from a statistical point of view, we found that the model does not significantly improve the residuals around the \ion{O}{7} He$\alpha$ line as presented in Figure~\ref{frcompare}.
We therefore conclude that RS is not the main cause for the observed anomalous \fr \ in \g.

CX is another possibility to account for the high \fr \  because the forbidden line intensity becomes relatively dominant by this process.
We added the CX model \citep{gu2016} to the above model,  ``(NEI$+$CX)$\times$abs'', coupling all the abundances and plasma temperature with those of the NEI component.
We varied the volume emission measure $\rm{VEM}_{\rm{CX}}$ $(=n_{\rm{p}} n_{\rm{NH}} V_{\rm{CX}}$, where $n_{\rm{NH}}$ and $V_{\rm{CX}}$ are the neutral hydrogen density and the effective interaction volume, respectively) and the collision velocity $v_{\rm{col}}$. 
We tried two collision cases, which are available in the CX model: 
a simpler case, which assumes a single collision for each ion, results in a physically unacceptable fit, in which $v_{\rm{col}}$ is much higher than that expected from a middle-aged SNR and  $kT_{\rm{e}}$ is too low to emit X-rays.
The other case assumes that one ion repeatedly captures electrons until it becomes neutral.
It seems more realistic assumption if SNR shocks are propagating into dense clouds.
We applied this ``multiple collision'' model and confirmed that the spectra are well explained with physically reasonable parameters.
The best-fit models  are shown in Figure~\ref{nwpl}.  
The parameters are summarized in Table~\ref{tab:par}.
The continuum is well reproduced by the NEI component, while the CX emission enhances  the lines, in particular the \oseven \ forbidden line, as indicated in Figure~\ref{frcompare}.

\section{Discussion} \label{sec:dis}
\subsection{Interaction with Molecular Clouds}\label{sec:cloud}
Through the analysis described in section~\ref{sec:ana}, we found evidence of CX in both NW and SE shells.
The detection of CX implies that the shock-heated plasma is interacting with a cold neutral gas population.
The estimated collision velocities $v_{\rm{col}}$ are 400~km~s$^{-1}$ and 200~km~s$^{-1}$ for the NW and SE shells, respectively (Table~\ref{tab:par}).
On the other hand, the shock velocity of \g \ is $v_{\rm{sh}}\sim600$~km~s$^{-1}$, which is derived from the best-fit value of $kT_{\rm{e}}$, using the Rankine-Hugoniot relation under a thermal equilibrium condition $v_{\rm{sh}}^{2}=16kT_{\rm{e}}$/$3\mu m_{\rm{H}}$, where $\mu$ is the mean molecular weight (=0.6) and $m_{\rm{H}}$ is the hydrogen mass.
Given that a plasma velocity $v$ behind the shock equals to $3/4 v_{\rm{sh}}\sim450$~km~s$^{-1}$, we found a potential discrepancy between the estimated collision velocities and the plasma velocity: $v>v_{\rm{col}}$.
A similar result was obtained by \citet{uchida2019} for an outermost rim of the Cygnus Loop ($v_{\rm{sh}}\sim300$~km~s$^{-1}$; $v_{\rm{col}}<50$~km~s$^{-1}$).
Both results imply that ions with a bulk velocity behind the shock is slowing down and undergoes collisions with a dense ambient medium, which may hint at the presence of CX in these SNRs.

To investigate the distribution of neutral molecular and atomic gas around \g, we analyzed $^{12}$CO($J$=1--0) and \ion{H}{1} data obtained with the NANTEN 4-m millimeter/submillimeter radio telescope \citep{mizuno2004} and the Australia Telescope Compact Array and Parkes 64-m radio telescope \citep{mcc2005}, respectively.

\begin{figure*}[ht!]
\centering
\includegraphics[width=150 mm]{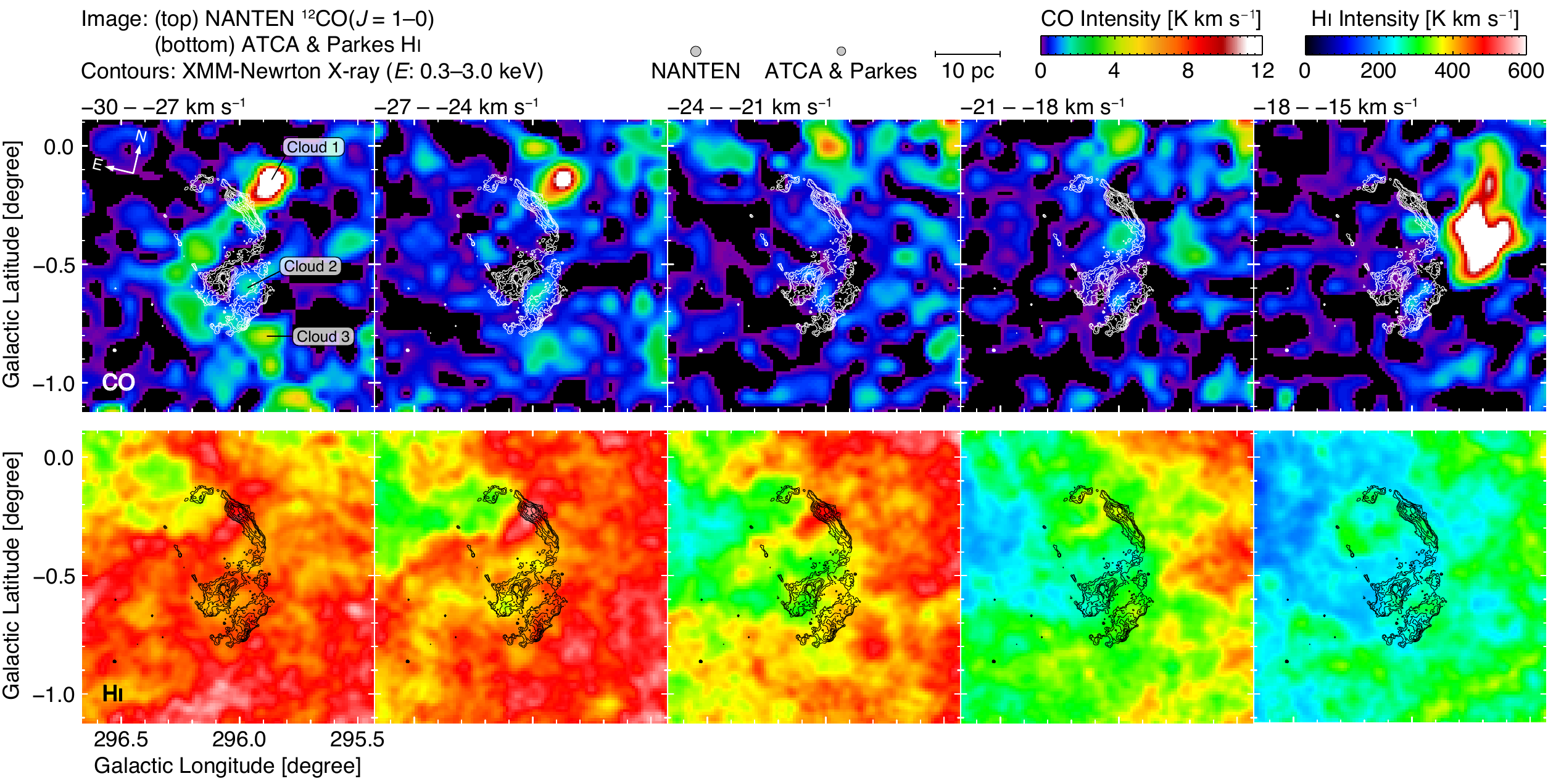}
\caption{Velocity channel distributions of the NANTEN $^{12}$CO($J$~=~1--0) intensities (top panels) and ATCA \& Parkes \ion{H}{1} intensities (bottom panels), superposed on the XMM-Newton X-ray intensity contours in the energy band of 0.3--3.0~keV. 
Each panel shows CO and \ion{H}{1} intensity distributions integrated every 3~km s$^{-1}$ in a velocity range from $-30$ to $-15$~km s$^{-1}$. 
The contour levels are 70, 100, 190, 340, 550, and 820 counts s$^{-1}$ degree$^{-2}$. The molecular clouds 1--3 discussed in Section~\ref{sec:cloud} are also indicated.}
\label{hi_co_channel}
\end{figure*}

Figure~\ref{hi_co_channel} shows the velocity channel distributions of CO and  \ion{H}{1} limited to the velocity range in which the CO and \ion{H}{1} clouds show good spatial correlation with the X-ray shell.
We found three molecular clouds in the NW and SE shells (hereafter Cloud 1 and Cloud 2, respectively) and toward the southwestern rim (hereafter, Cloud 3), which are likely associated with \g. 
The distribution of \ion{H}{1} gas at $V_{\mathrm{LSR}}=-24$~km~s$^{-1}$ to $V_{\mathrm{LSR}}=-21$~km~s$^{-1}$ is also well correlated with the X-ray shell in the west.
These results imply a presence of dense gas toward the west and are also consistent with the faint X-ray emission  in the east.

\begin{figure*}[ht!]
\centering
\includegraphics[width=150 mm]{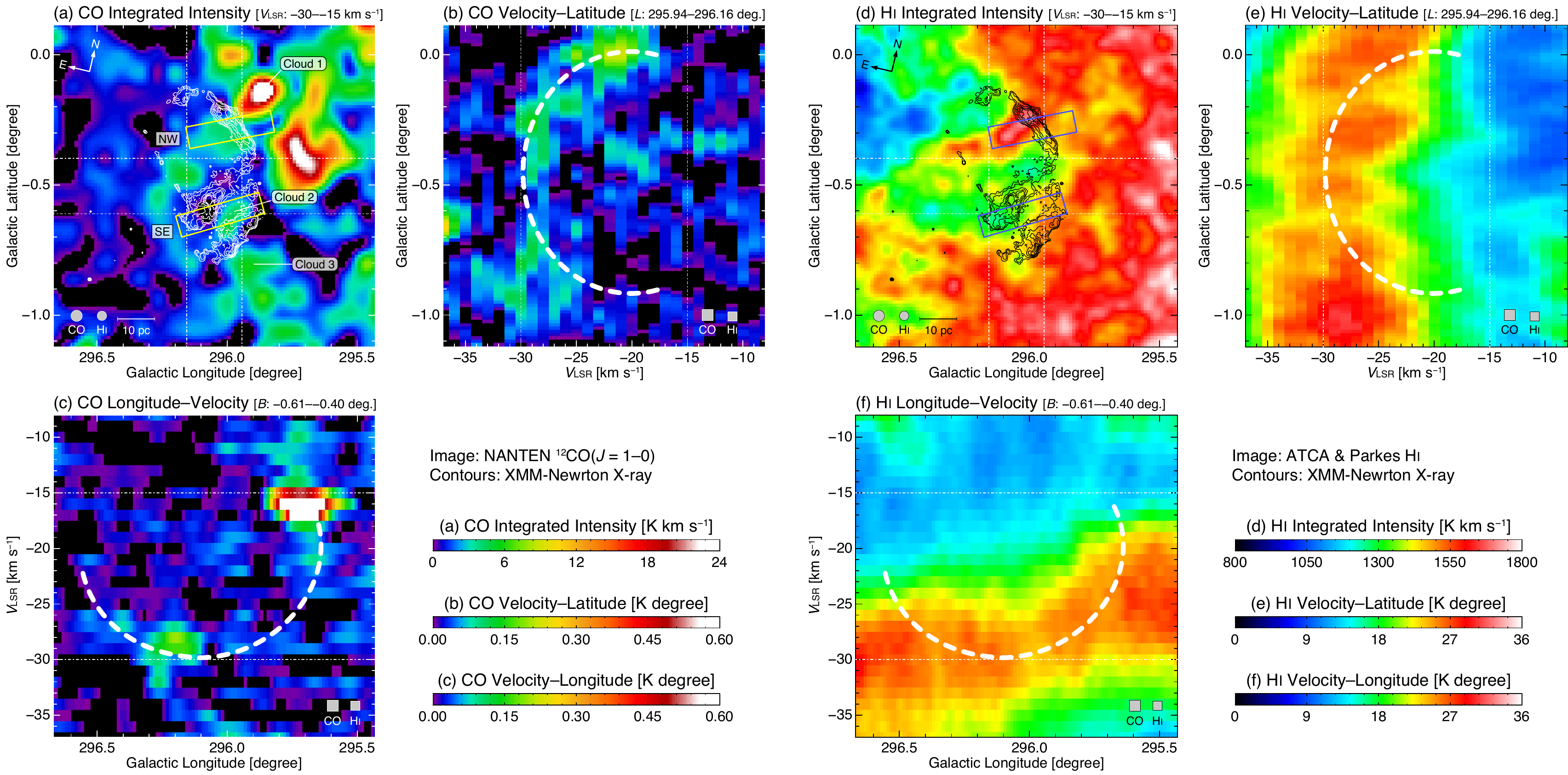}
\caption{Velocity integrated intensity maps and position--velocity (p--v) diagrams of CO (left panels) and \ion{H}{1} (right panels). 
The integration range is from $-$30 to $-15$~km s$^{-1}$ in velocity for each intensity map (a and d); from 295\fdg94 to 296\fdg15 in Galactic Longitude for each velocity--latitude diagram (b and e); and from $-0\fdg61$ to $-0\fdg40$ in Galactic Latitude for each longitude--velocity diagram (c and f). 
The superposed contours are the same as shown in Figure~\ref{hi_co_channel}. 
The rectangles, NW and SE, indicate the regions used for extracting RGS spectra. 
Clouds~1--3 discussed in Section~\ref{sec:cloud} are also indicated. 
The dashed curve in each p--v diagram represents the boundaries of the CO and \ion{H}{1} cavities.}
\label{hi_co_pv}
\end{figure*}

Figures~\ref{hi_co_pv} (a) and (d) show the velocity integrated intensity maps of CO and \ion{H}{1} at $V_{\mathrm{LSR}}=-30$~km~s$^{-1}$ to $V_{\mathrm{LSR}}=-15$~km~s$^{-1}$, respectively.
The spatial relation between the X-ray shell and the neutral gas is clearly revealed in each map.
Figures~\ref{hi_co_pv} (b), (c) and (e), (f) show the position--velocity (p--v) diagrams of CO and \ion{H}{1}, respectively.
We found a cavity-like distribution along with CO and \ion{H}{1} clouds for each p--v diagram. 
The  distribution and the spatial shift of \ion{H}{1}  suggest that  the CO and \ion{H}{1} clouds are expanding with a velocity of $\sim 10$~km~s$^{-1}$ and a systemic velocity of $-20 \pm 4$~km~s$^{-1}$.
Since the spatial extent of the CO/\ion{H}{1} emissions  is roughly consistent with the apparent diameter of the X-ray shell, we conclude that the cavity was formed by shock waves and/or a stellar wind from the progenitor of \g \ \citep[cf.][]{koo1990}.
These results confirm that Clouds~1--3 and \ion{H}{1} gas at $V_{\mathrm{LSR}}=-30$~km~s$^{-1}$ to $V_{\mathrm{LSR}}=-15$~km~s$^{-1}$ are associated with \g, while further follow-up observations would be needed to obtain more detailed information such as the intensity ratio of CO $J$ = 3--2/1--0, their line broadening, and a sub-pc-scale spatial resolution \citep[e.g.,][]{seta1998, sano2021}.
By adopting the standard Galactic rotation curve model with the IAU recommended values of $R_{0} = 8.5$~kpc and $\Theta_{0} = 220$~km~s$^{-1}$ \citep{kerr1986, brand1993}, we estimate the distance to the SNR to be $\sim2.1^{+0.9}_{-0.6}$~kpc, which is consistent with a previous analysis \citep[2--5~kpc;][]{castro2011} and gives more reliable result.

From the radio observations, we speculate that the \ion{H}{1} gas is globally associated with \g \ in the west and contributes to the CX emission, whereas the dense clouds such as Clouds~1 and 2 are likely outside the shells.
We calculate the hydrogen number density of \ion{H}{1} gas near the clouds as $\sim 200$~cm$^{-3}$ for NW  and 160~cm$^{-3}$ for SE by assuming an optically thin emission \citep[e.g.,][]{dicky1990} and the depth of the \ion{H}{1} gas to be $\sim$5~pc.
Basic physical properties for each molecular cloud are listed in Table~\ref{tab:mc}, for which we adopted the SNR distance of 2.1~kpc and a CO-to-H$_{2}$ conversion factor of $2.0 \times 10^{20}~\mathrm{cm^{-2} (K~km~s^{-1})^{-1}}$ \citep[e.g.,][]{bolatto2013}.

Since a non-negligible fraction of hydrogen atoms in molecular clouds undergo collisional ionization, a number of hydrogen atoms that can cause CX is limited.
Relative probability of the CX and electron collisional ionization depends on  the collision velocity and plasma temperature \citep{lallement2004}.
If the collision velocity is high ($\sim1000$~km~s$^{-1}$), hot ions can stream deeply into a molecular cloud, resulting in a relatively high probability of the CX.
In a case of a high temperature plasma ($\sim1$~keV), collisional excitation is dominant and neutral hydrogen atoms are mostly ionized.
Applying $v_{\rm{col}}$ and  $kT_{\rm{e}}$ of the best-fit (NEI$+$CX) $\times$ abs models (Table~\ref{tab:par}) to the discussion by \citet{lallement2004} (see their Figure~1), we assume that $\sim30$\% (for NW) and $\sim20$\% (for SE) of the neutral gas densities  contribute to the CX process before the collisional ionization.
As a result, we obtain neutral hydrogen densities $n_{\rm{NH}}$ to be  $\sim 60$~cm$^{-3}$ and $\sim 32$~cm$^{-3}$ for NW and SE, respectively.

\setcounter{table}{3}
\begin{deluxetable*}{lcccccccc}[]
\tablecaption{Physical properties of molecular clouds likley associated with \g}
\tablehead{\\
\multicolumn{1}{c}{Cloud name} & $\alpha_{\mathrm{J2000}}$ & $\delta_{\mathrm{J2000}}$ & $T_{\mathrm{mb}} $ & $V_{\mathrm{LSR}}$ & $\Delta V$ & Size &  Mass & $n_\mathrm{H}(\mathrm{H_2})$\\
& ($^{\mathrm{h}}$ $^{\mathrm{m}}$ $^{\mathrm{s}}$) & ($^{\circ}$ $\arcmin$ $\arcsec$) & (K) & \scalebox{0.9}[1]{(km $\mathrm{s^{-1}}$)} & \scalebox{0.9}[1]{(km $\mathrm{s^{-1}}$)} & (pc) &  ($M_\sun $) & (cm$^{-3}$)\\
\multicolumn{1}{c}{(1)} & (2) & (3) & (4) & (5) & (6) & (7) & (8) & (9)}
\startdata
Cloud~1 ................. & 11 50 21.15 & $-$62 10 00.6 & $7.35 \pm 0.14$ & $-27.54 \pm 0.04$ & $4.28 \pm 0.09$ & 3.8 & 1200 & 1600\\
Cloud~2 ................. & 11 50 07.41 & $-$62 38 56.7 & $0.56 \pm 0.13$ & $-27.28 \pm 0.55$ & $4.84 \pm 1.37$ & 5.0 & \phantom{0}230 & \phantom{0}140\\
Cloud~3 ................. & 11 49 05.94 & $-$62 48 48.4 & $1.61 \pm 0.14$ & $-29.54 \pm 0.17$ & $4.12 \pm 0.41$ & 5.8 & \phantom{0}590 & \phantom{0}230\\
\enddata
\tablecomments{Col.~(1): Cloud name. Cols. (2--9): Observed properties of clouds obtained by a single Gaussian fitting with $^{12}$CO($J$~=~1--0) line emission. Cols.~(2)--(3): Position of clouds in the equatorial coordinate. Col.~(4): Peak radiation temperature. Col.~(5): Central velocities of CO spectra. Col.~(6): Full Width Half Maximum (FWHM) linewidths of CO spectra $\Delta V$. Col.~(7): Diameters of clouds defined as $(A / \pi)^{0.5} \times 2$, where $A$ is the surface area of each cloud surrounded by a contour of the half level of the maximum integrated intensity. Col. (8): Masses of clouds derived by an equation of $N(\mathrm{H_2}) / W(\mathrm{CO}) = 2.0 \times 10^{20}$ (K km s$^{-1}$)$^{-1}$ cm$^{-2}$, where $N(\mathrm{H_2})$ is the molecular hydrogen column density and $W$(CO) is the integrated intensity of $^{12}$CO($J$~=~1--0) \citep{bolatto2013}. Col.~(9): Hydrogen number densities of clouds $n_\mathrm{H}(\mathrm{H_2})$. The errors of mass and $n_\mathrm{H}(\mathrm{H_2})$ are $\sim$30\% by assuming the same error in the CO-to-H$_2$ conversion factor \citep[cf.][]{bolatto2013}.}
\label{tab:mc}
\end{deluxetable*}

\subsection{Emitting Region of CX}\label{sec:cx}
Previous theoretical calculations predict that CX occurs in a thin layer at the outermost edge of a SNR blast wave \citep[e.g.,][]{lallement2004}.
According to an estimation for a  spherical SNR, CX dominates over thermal emission in the outer layer with a ``collision parameter'' of $p\sim0.99$ \citep{lallement2004}; 
when the CX-emitting region has a thickness of $\Delta r$ from the shock front, $p$ is expressed as 
\begin{eqnarray}\label{p}
p=\frac{r_{s(out)} - \Delta r}{r_{s(out)}},
\end{eqnarray}
where the outer radius of  swept-up ISM is $r_{s(out)}$.
In the case of  $p = 0.99$, the outer layer with a thickness of 1\% of the shock radius is CX dominant.
Following this result, \citet{katsuda2011} claimed that the expected CX emission measure in the Cygnus Loop is consistent with  the theoretical value.
However, there was a discrepancy by a factor of 2--6.
A more accurate value of $p$ is required but a detailed measurement has not been obtained so far.

\begin{figure}[ht!]
\centering
\includegraphics[clip,width=80 mm]{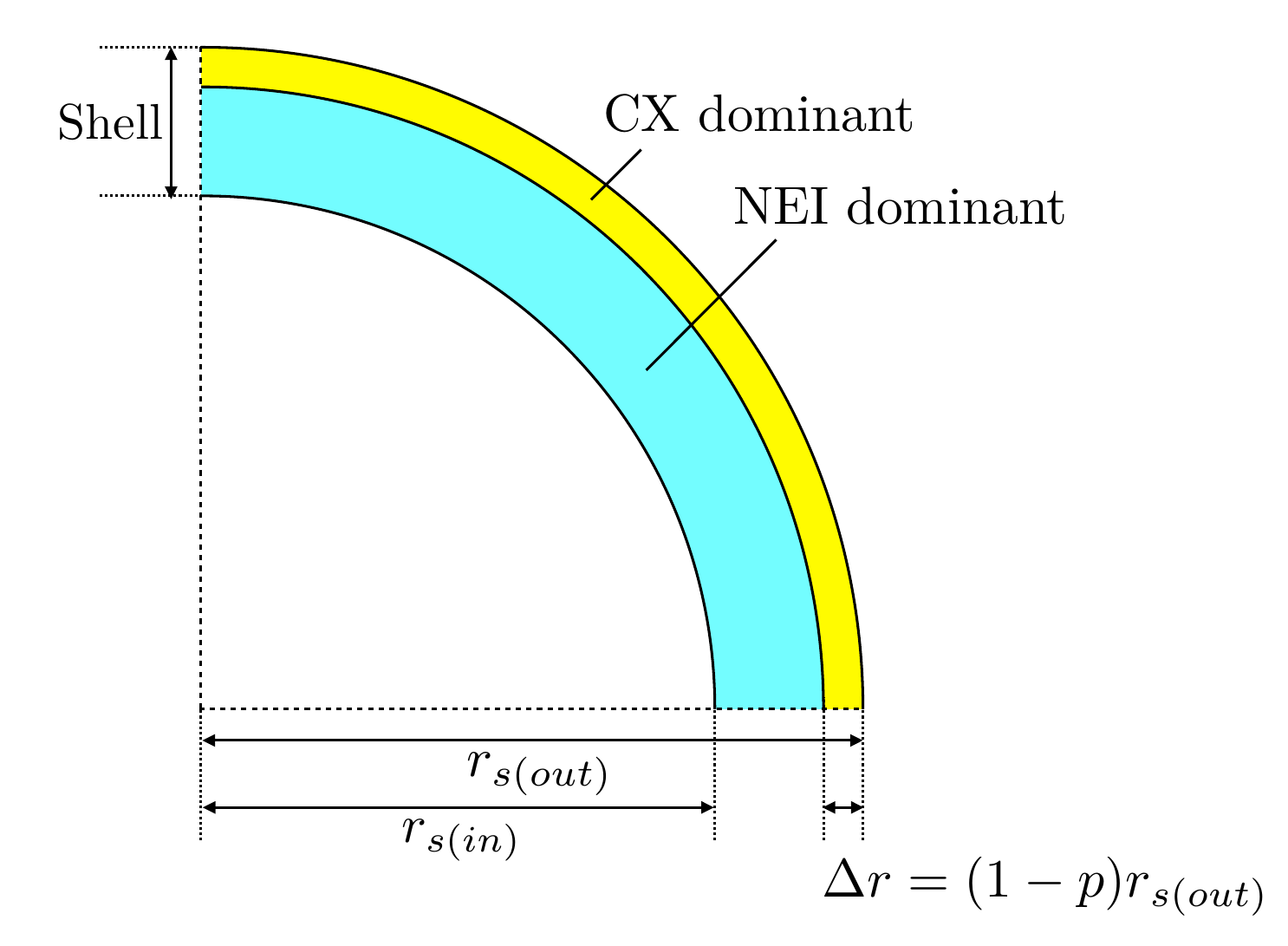}
\caption{Geometry of shells we assumed for the calculation described in the text. The cyan and yellow  represent NEI- and CX-dominant regions, respectively.}
\label{geo}
\end{figure}

We can estimate $p$ for the NW and SE shells since we obtained $n_{\rm{NH}}$ of the interacting molecular clouds in the CX-emitting regions.
As illustrated in Figure~\ref{geo}, we  assume a part of a spherical shell between the outer ($r_{s(out)}$) and inner ($r_{s(in)}$) radii.
The ratio of the emitting volume of the CX ($V_{\rm{CX}}$) to NEI ($V_\mathrm{NEI}$) is described with a collision parameter $p$ as follows:
\begin{eqnarray}\label{p}
\frac{V_{\rm{CX}}}{V_\mathrm{NEI}} = \frac{r_{s(out)}^{3} - (r_{s(out)} - \Delta r)^{3}}{r_{s(out)}^{3} - r_{s(in)}^{3}} = \frac{1 - p^{3}}{1 - (\frac{r_{s(in)}}{r_{s(out)}})^{3}}.
\end{eqnarray}
The ratio of the volume emission measure of the CX component to that of the NEI component is
\begin{eqnarray}\label{normratio}
\frac{\rm{VEM}_{\rm{CX}}}{\rm{VEM}_{\rm{NEI}}} = \frac{n_{\rm{p}} n_{\rm{NH}} V_{\rm{CX}}}{n_{\rm{e}} n_{\rm{p}} V_{\mathrm{NEI}}} \simeq \frac{n_{\rm{NH}} V_{\rm{CX}}}{1.2 n_{\rm{p}} V_\mathrm{NEI}},
\end{eqnarray}
where we assume the solar metallicity, i.e., $n_{\rm{e}}\simeq1.2n_{\rm{p}}$.

We get $V_{\mathrm{NEI}} \sim 5 \times 10^{56}~\mathrm{cm^{3}}$, assuming from the X-ray morphology that the NW rim is a part of a spherical shell with a thickness of 1~pc and a radius of 7~pc: $r_{s(out)}=7$~pc and $r_{s(in)}=6$~pc.
Since the best-fit value of $\rm{VEM}_{\rm{NEI}}$ is $1.0^{+0.3}_{-0.1} \times 10^{56}$~cm$^{-3}$ ($= 1.2 n^{2}_{\rm{p}} V_\mathrm{NEI}$), we derive $n_{\rm{p}} = 0.4\ \mathrm{cm^{-3}}$ for NW.
Substituting these values into Equation~\ref{normratio}, we obtain $V_{\rm{CX}}$/$V_{\mathrm{NEI}}$=$9.0 \times 10^{-3}$. 
From Equation~\ref{p}, the collision parameter in the NW shell is accordingly estimated to be $p=0.998$.
The same calculation for the SE shell gives, we derive $V_{\rm{CX}}$/$V_{\mathrm{NEI}}=9.2 \times 10^{-3}$ and $p=0.997$, where we assume  $r_{s(out)}=4$~pc, $r_{s(in)}=2$~pc and $V_{\mathrm{NEI}} \sim 2 \times 10^{57}$~cm$^{3}$.

While \citet{lallement2004} estimated that $\sim1\%$ of the shock radius is CX dominant based on the observation of the entire region of DEM~L71, our estimate from  partial shells of the nearby SNR \g \ provides more accurate measurement of $p$ than before.
The obtained values  are $p=0.998$ and $p=0.997$ for NW and SE, respectively, which means 0.2--0.3~\% of the shock radius is CX dominant.
Our result clearly indicates that  CX is  occurring at the outermost edge of the shock front.
This is consistent with  a general picture expected in SNRs.
Similar anomalous \frs \ have also been reported recently in other SNRs.
In these examples, the RS effect would be more preferable than CX if $p$ is much smaller (or $V_{\rm{CX}}$/$V_{\mathrm{NEI}}$ is much larger) than the above value \citep{suzuki2020}.

\subsection{Future works}\label{sec:future}
It is expected that different conditions such as the shock velocity, ambient density, and abundances lead to  different line ratios of CX \citep{gu2016}.
In this context, \g \ is the best case that enables us to study various locations within the same object.
As indicated in Figure~\ref{hi_co_channel}, another region where we predict to detect CX is the southernmost shell interacting with Cloud~3.
Since this shell is farthest from the SNR center, a higher forward shock velocity is expected, which may result in a different line ratio from those in the other shells.
In contrast, the northernmost edge of the NW shell can be interpreted as unlikely that the CX is occurring, because this region has little correlation with the $^{12}$CO emission.
These regions have a similar spatial width to the NW and SE shells and therefore future observations with the RGS will unveil the presence of CX almost entirely in \g.

If we assume our result can be applied to other nearby SNRs, it is possible to identify regions where CX is dominantly occurring.
As pointed out in several studies \citep{katsuda2012, roberts2015, uchida2019}, the Cygnus Loop is one of the plausible candidates for detecting CX.
Given that the Cygnus  Loop has a spherical shell with an apparent diameter of $3^\circ$, we expect that a pure CX emission can be detected in a thickness of $\sim15\arcsec$ from the shock front.
If this is the case, XRISM Resolve \citep{tashiro2018}, which has a pixel size of $\sim30\arcsec$, is likely to detect a CX-dominant spectrum in the outermost edge of the remnant.
Athena X-ray Integral Field Unit \citep[X-IFU;][]{athena} and Lynx X-ray Microcalorimeter \citep[LXM;][]{lynx} will provide us with a larger sample of SNRs, in which we can extract only  CX X-ray emission.

\section{Conclusion} \label{sec:conc}
We performed high-resolution spectral analysis of two shell regions (NW and SE) of \g \ with the RGS onboard XMM-Newton and discovered that the \fr s of  \oseven \ He$\alpha$ are significantly higher  in both the shells than those expected from  a normal NEI model.
This is the first time that such anomalous \oseven \ He$\alpha$ line is detected from multiple locations in a single remnant, although similar results have been obtained from observations of other SNRs \citep[e.g.,][]{uchida2019}.
The spectra of the NW and SE shells are well represented by incorporating a CX model with the previously reported NEI model.
We also revealed that \g \ is interacting with molecular clouds in several locations including NW and SE, which is based on our $^{12}$CO($J$ = 1--0) and \ion{H}{1} observation around the remnant.
These results strongly support  a picture that CX is occurring in the region where high \fr s are detected.
These molecular clouds have velocities of $V_{\mathrm{LSR}}=-30$~km~s$^{-1}$ to $-15$~km~s$^{-1}$, from which we estimated the distance to \g \ to be 2.1~kpc. 
Assuming a spherical geometry, we calculated  the contribution of the CX emission in the shells and concluded that CX is dominant in a region with a thickness of 0.2--0.3\%  of the shock radius (collision parameter $p\sim0.997$--0.998).
This is consistent with a previous theoretical calculation by \citet{lallement2004}.
The obtained value of $p$ will be useful for future observations with high-resolution spectroscopy such as XRISM and Athena to quantitatively evaluate a CX-dominant region in nearby SNRs.

\begin{acknowledgments}
The NANTEN project is based on a mutual agreement between Nagoya University and the Carnegie Institution of Washington (CIW).
We greatly appreciate the hospitality of all the staff members of the Las Campanas Observatory of CIW.
We are thankful to many Japanese public donors and companies who contributed to the realization of the project. 
This work is supported by JSPS KAKENHI Scientific Research Grant Numbers JP19K03915 (H.U.), JP19H01936 (T.T.), JP20KK0309 (H.S.), JP21H01136 (H.S.) and JP21H04493 (T.G.T and T.T.).
\end{acknowledgments}

\bibliography{sample631.bib}{}
\bibliographystyle{aasjournal}

\end{document}